\documentclass[10pt]{article}
\usepackage{graphicx}
\usepackage{amsmath}
\usepackage{amssymb}
\usepackage{caption2}
\begin{document}
\begin{center}
{\large {\bf \sc{  Analysis of the hidden-charm tetraquark molecule mass spectrum  with the QCD sum rules }}} \\[2mm]
Zhi-Gang  Wang \footnote{E-mail: zgwang@aliyun.com.  }     \\
 Department of Physics, North China Electric Power University, Baoding 071003, P. R. China
\end{center}

\begin{abstract}
In this article,  we  investigate the mass spectrum of the ground state hidden-charm tetraquark molecular states without strange, with strange and with hidden-strange via the QCD sum rules in a comprehensive way and revisit the assignments of the $X$, $Y$, $Z$ states in the  scenario of tetraquark  molecular states  consistently  based on the QCD sum rules.
 \end{abstract}



\section{Introduction}
In 2003, the  Belle collaboration   observed   a narrow charmonium-like state $X(3872)$ in the $\pi^+ \pi^- J/\psi$ invariant  mass spectrum \cite{X3872-2003}, which cannot be accommodated in the traditional  two quark model comfortably. Up to now, about twenty   charmonium-like (and bottomonium-like) states were observed  by the  BaBar, Belle, BESIII, CDF, CMS, D0, LHCb collaborations \cite{PDG},   which cannot be accommodated in the traditional  two quark model in one way or the other, and are excellent candidates for the tetraquark states and tetraquark molecular states. They are usually named as the exotic $X$, $Y$ and $Z$ states.

 In 2006, R. D. Matheus et al took the $X(3872) $ as the  diquark-antidiquark type  tetraquark state with the quantum numbers $J^{PC}=1^{++}$, and studied  its mass with the QCD sum rules   by carrying out the operator product expansion up to the vacuum condensates  of dimension 8 \cite{Narison-3872}.  Thereafter   the QCD sum rules became a powerful theoretical approach in studying the masses and decay widths of the $X$, $Y$ and $Z$ states to diagnose their natures, irrespective of the  hidden-charm (or hidden-bottom) tetraquark states \cite{Azizi-Z4430-tetra-6,Azizi-Zcs3985-tetra-6,SLZhu-tetra-8,SLZhu-tetra-AV-8,CFQiao-tetra-8,HXChen-tetra-8,CFQiao-Zcs3985-tetra-mol-8,
WangHuangtao-2014-PRD,Wang-tetra-formula,WZG-Z10610-tetra-10,WZG-Y4500-tetra-10,WZG-Y4600-tetra-10,
WZG-V-tetra-10,Azizi-Z10610-tetra-10,WZG-Y4140-tetra-10,WZG-Y4274-tetra-10,
WZG-hidden-bb-tetra-10,WZG-hidden-cc-tetra-10,WZG-Z3900-two-particle-10,WZG-Zcs3985-tetra-10,WZG-Z4600-tetra-10} or hadronic molecular states \cite{Nielsen-mole-Zcs-6,JRZhang-mole-6,JRZhang-Z10610-mol-6,Nielsen-X3872-mol-cc-8,Narison-Z4430-mol-8,Nielsen-Y4260-mol-8,HXChen-Zcs3985-mol-8,YLLiu-Zcs3985-mol-8,WZG-Y4140-mole-10,
WZG-Y4660-mole-10,WZG-Z3900-mole-10,WZG-Z4020-mole-10, WZG-Y4260-mol-10-CPC,WZG-DZY-mole-10}.

In the QCD sum rules, we usually construct  the color-antitriplet-color-triplet ($ \bar{\bf3}_c{\bf 3}_c$)-type, in other words, the diquark-antidiquark type \cite{Narison-3872,Azizi-Z4430-tetra-6,Azizi-Zcs3985-tetra-6,SLZhu-tetra-8,SLZhu-tetra-AV-8,CFQiao-tetra-8,HXChen-tetra-8,CFQiao-Zcs3985-tetra-mol-8,
WangHuangtao-2014-PRD,Wang-tetra-formula,WZG-Z10610-tetra-10,WZG-Y4500-tetra-10,WZG-Y4600-tetra-10,
WZG-V-tetra-10,Azizi-Z10610-tetra-10,WZG-Y4140-tetra-10,WZG-Y4274-tetra-10,
WZG-hidden-bb-tetra-10,WZG-hidden-cc-tetra-10,WZG-Z3900-two-particle-10,WZG-Zcs3985-tetra-10,WZG-Z4600-tetra-10},
color-sextet-color-antisextet (${\bf 6}_c\bar{\bf 6}_c$)-type \cite{SLZhu-tetra-8,SLZhu-tetra-AV-8,HJLee-PRD-a980-66},
 color-singlet-color-singlet (${\bf 1}_c{\bf 1}_c$)-type \cite{Nielsen-mole-Zcs-6,JRZhang-mole-6,JRZhang-Z10610-mol-6,Nielsen-X3872-mol-cc-8,Narison-Z4430-mol-8,Nielsen-Y4260-mol-8,HXChen-Zcs3985-mol-8,YLLiu-Zcs3985-mol-8,WZG-Y4140-mole-10,WZG-Y4660-mole-10,WZG-Z3900-mole-10,WZG-Z4020-mole-10,
 WZG-Y4260-mol-10-CPC,WZG-DZY-mole-10}
 and color-octet-color-octet (${\bf 8}_c{\bf 8}_c$)-type \cite{WZG-Z3900-mole-10,CFQiao-88-tetra-8,WZG-Z4200-88-tetra-10,WZG-NPA-Y2175-88} local four-quark current operators to interpolate  the tetraquark states. In the literatures, we usually refer the $ \bar{\bf3}_c{\bf 3}_c$-type and ${\bf 1}_c{\bf 1}_c$-type tetraquark states as the tetraquark states and tetraquark (or hadronic) molecular states, respectively. As we choose the local current operators, it is better to refer the corresponding tetraquark states as the $ \bar{\bf3}_c{\bf 3}_c$-type, ${\bf 6}_c\bar{\bf 6}_c$-type, ${\bf 1}_c{\bf 1}_c$-type and $ {\bf8}_c{\bf 8}_c$-type tetraquark states, respectively.

 The diquark-antidiquark type local four-quark  current   with definite  quantum numbers couples potentially  to a definite compact tetraquark state, though this  local current  can be re-arranged into a special superposition of  a series of color-singlet-color-singlet type  currents through Fierz transformation, which couple potentially  to the tetraquark molecular states or  two-meson scattering  states  with the same quantum numbers, as there are no barriers to frustrate the rearrangements in the coordinate space.
The diquark-antidiquark type tetraquark states can be taken as a special superposition of  a series of color-singlet-color-singlet compact molecular states and embody the net effects, and vise versa. In the QCD sum rules, the local currents require that the tetraquark molecular states have the average spatial sizes $\sqrt{\langle r^2\rangle}$ of the same magnitudes as the conventional mesons and baryons, and  they are also compact objects, just like the diquark-antidiquark type tetraquark states.

 In carrying out the operator product expansion, we usually   truncate  the vacuum condensates up to dimension $6$ \cite{Azizi-Z4430-tetra-6,Azizi-Zcs3985-tetra-6,Nielsen-mole-Zcs-6,JRZhang-mole-6,JRZhang-Z10610-mol-6}, $8$  \cite{Narison-3872,SLZhu-tetra-8,SLZhu-tetra-AV-8,CFQiao-tetra-8,HXChen-tetra-8,CFQiao-Zcs3985-tetra-mol-8,Nielsen-X3872-mol-cc-8,
 Narison-Z4430-mol-8,Nielsen-Y4260-mol-8,HXChen-Zcs3985-mol-8,YLLiu-Zcs3985-mol-8,
 CFQiao-88-tetra-8}, or $10$ \cite{WangHuangtao-2014-PRD,Wang-tetra-formula,WZG-Z10610-tetra-10,WZG-Y4500-tetra-10,WZG-Y4600-tetra-10,
WZG-V-tetra-10,Azizi-Z10610-tetra-10,WZG-Y4140-tetra-10,WZG-Y4274-tetra-10,
WZG-hidden-bb-tetra-10,WZG-hidden-cc-tetra-10,WZG-Z3900-two-particle-10,WZG-Zcs3985-tetra-10,WZG-Z4600-tetra-10,WZG-Y4140-mole-10,WZG-Y4660-mole-10,WZG-Z3900-mole-10,WZG-Z4020-mole-10,
 WZG-Y4260-mol-10-CPC,WZG-DZY-mole-10,WZG-Z4200-88-tetra-10}. Although the convergence of the operator product expansion requires that the contributions of the vacuum condensates of dimension $10$, which are of the highest dimension,  play a tiny important role in the Borel windows, the highest vacuum condensates are companied with the factors $\frac{1}{T^2}$, $\frac{1}{T^4}$, $\frac{1}{T^6}$, or $\frac{1}{T^8}$, and play an important role in determining the Borel windows so as to extract the reliable  tetraquark (molecule) masses, we should take them into account in a consistent way, where the $T^2$ is the Borel parameter.

For the exotic states  $X_c(3872)$, $Z_c(3900)$, $Z_{cs}(3985)$, $Z_c(4020)$, $Y(4390)$, $Y(4660)$, $Z_b(10610)$, $Z_b(10650)$,  we can reproduce the experimental values of the masses both in the  scenarios  of tetraquark  states \cite{Narison-3872,Azizi-Zcs3985-tetra-6,CFQiao-Zcs3985-tetra-mol-8,WangHuangtao-2014-PRD,Wang-tetra-formula,WZG-Z10610-tetra-10,WZG-Y4600-tetra-10,WZG-V-tetra-10,
Azizi-Z10610-tetra-10,WZG-hidden-bb-tetra-10,WZG-hidden-cc-tetra-10,WZG-Z3900-two-particle-10,WZG-Zcs3985-tetra-10,WZG-Z4600-tetra-10}  and tetraquark molecular states \cite{CFQiao-Zcs3985-tetra-mol-8,Nielsen-mole-Zcs-6,JRZhang-Z10610-mol-6,Nielsen-X3872-mol-cc-8,HXChen-Zcs3985-mol-8,YLLiu-Zcs3985-mol-8,
WZG-Y4660-mole-10,WZG-Z3900-mole-10,WZG-Z4020-mole-10,WZG-Y4260-mol-10-CPC}.
While for the exotic states $X(3860)$, $X(4140)$, $X(4274)$, $Y(4220/4260)$,  $Y(4320/4360)$, $Z_c(4250)$, $X(4500)$, $X(4700)$,  we can reproduce the experimental values of the masses only  in the  scenario of tetraquark  states \cite{WZG-Y4500-tetra-10,WZG-V-tetra-10,WZG-Y4140-tetra-10,WZG-Y4274-tetra-10,WZG-hidden-cc-tetra-10}.
In the framework of the QCD sum rules, the tetraquark  scenario can accommodate more exotic states than the tetraquark  molecular scenario.
However, we should not be  pessimistic about the tetraquark molecular scenario, as the works on the tetraquark states are also overwhelming, a comprehensive analysis based on the tetraquark molecular scenario is still needed.

In the QCD sum rules for the multiquark states, it is difficult (or not easy) to satisfy the pole dominance at the hadron side.
In Ref.\cite{WangHuangtao-2014-PRD}, we investigate  the $X_c(3872)$ and $Z_c(3900)$ as   the diquark-antidiquark type  axialvector tetraquark states with the QCD sum rules, and  explore the energy scale dependence of the QCD sum rules for the hidden-charm   tetraquark states   for the first time. In  the subsequent works,  we suggest an energy scale formula,
\begin{eqnarray}
\mu&=&\sqrt{M^2_{X/Y/Z}-(2{\mathbb{M}}_Q)^2} \, ,
 \end{eqnarray}
 with the effective heavy  quark masses ${\mathbb{M}}_Q$ to determine the optimal energy scales of the QCD spectral densities for  the hidden-charm and hidden-bottom tetraquark states and tetraquark molecular states \cite{Wang-tetra-formula,WZG-Z10610-tetra-10,WZG-V-tetra-10,WZG-Z3900-mole-10,WZG-Z4020-mole-10}. The energy scale formula can enhance the pole contributions at the hadron side remarkably, and also improve the convergent behavior of the operator product expansion at the QCD side remarkably.
In Refs.\cite{WZG-hidden-bb-tetra-10,WZG-hidden-cc-tetra-10}, we  investigate  the  mass spectrum of the ground state hidden-charm and hidden-bottom tetraquark states  with the QCD sum rules in a comprehensive way, and  revisit the assignments of  the $X(3860)$, $X(3872)$, $X(3915)$,  $X(3940)$, $X(4160)$, $Z_c(3900)$, $Z_c(4020)$, $Z_c(4050)$, $Z_c(4055)$, $Z_c(4100)$, $Z_c(4200)$, $Z_c(4250)$, $Z_c(4430)$, $Z_c(4600)$, $Z_b(10610)$ and $Z_b(10650)$ in the  scenario of tetraquark  states in a consistent way based on the QCD sum rules.

Now we extend our previous works \cite{WZG-Z3900-mole-10,WZG-Z4020-mole-10} to study the ground state mass spectrum of the hidden-charm tetraquark molecular states without strange, with strange and with hidden-strange via the QCD sum rules by carrying out the operator product expansion up to the vacuum condensates of dimension $10$ in a consistent way and  taking  account of the $SU(3)$ breaking effects also in a consistent way, and revisit the assignments of the $X$, $Y$ and $Z$ states in the scenario of the tetraquark molecular states to examine the exotic properties again. Furthermore, we obtain reliable predictions of the tetraquark molecule  mass spectrum, which can be confronted to the experimental data in the future   at the BESIII, LHCb, Belle II,  CEPC, FCC, ILC, and shed light on the nature of the exotic $X$, $Y$, $Z$ states. Compared to previous works \cite{CFQiao-Zcs3985-tetra-mol-8,Nielsen-mole-Zcs-6,JRZhang-mole-6,JRZhang-Z10610-mol-6,Nielsen-X3872-mol-cc-8,Narison-Z4430-mol-8,Nielsen-Y4260-mol-8}, we take  account of more vacuum condensates, obtain larger pole contributions and more flat Borel platforms,  and perform  a comprehensive analysis.

The article is arranged as follows:  we derive the QCD sum rules for the masses and pole residues  of  the  hidden-charm tetraquark molecular states in section 2; in section 3, we   present the numerical results and discussions; section 4 is reserved for our conclusion.

\section{QCD sum rules for  the  hidden-charm  tetraquark molecular states}

Firstly, let us  write down  the two-point correlation functions $\Pi(p)$, $\Pi_{\mu\nu}(p)$ and $\Pi_{\mu\nu\alpha\beta}(p)$  in the QCD sum rules,
\begin{eqnarray}\label{CF-Pi}
\Pi(p)&=&i\int d^4x e^{ip \cdot x} \langle0|T\Big\{J(x)J^{\dagger}(0)\Big\}|0\rangle \, ,\nonumber\\
\Pi_{\mu\nu}(p)&=&i\int d^4x e^{ip \cdot x} \langle0|T\Big\{J_\mu(x)J_{\nu}^{\dagger}(0)\Big\}|0\rangle \, ,\nonumber\\
\Pi_{\mu\nu\alpha\beta}(p)&=&i\int d^4x e^{ip \cdot x} \langle0|T\Big\{J_{\mu\nu}(x)J_{\alpha\beta}^{\dagger}(0)\Big\}|0\rangle \, ,
\end{eqnarray}
where the currents $J(x)=J_{D^*\bar{D}^*}(x)$, $J_{D^*\bar{D}_s^*}(x)$, $J_{D_s^*\bar{D}_s^*}(x)$,  $J_\mu(x)=J_{D\bar{D}^*,\pm,\mu}(x)$,
 $J_{D\bar{D}_s^*,\pm,\mu}(x)$,   $J_{D_s\bar{D}_s^*,\pm,\mu}(x)$,   $J_{\mu\nu}(x)=J_{\pm,\mu\nu}(x)$, $J_{\pm,\mu\nu}(x)=J_{D^*\bar{D}^*,\pm,\mu\nu}(x)$,
$J_{D^*\bar{D}_s^*,\pm,\mu\nu}(x)$, $J_{D_s^*\bar{D}_s^*,\pm,\mu\nu}(x)$,
\begin{eqnarray}
J_{D^*\bar{D}^*}(x)&=& \bar{q}(x)\gamma_\mu  c(x)\,   \bar{c}(x)\gamma^\mu  q(x) \, ,\nonumber \\
J_{D^*\bar{D}_s^*}(x)&=& \bar{q}(x)\gamma_\mu  c(x)\,   \bar{c}(x)\gamma^\mu  s(x) \, ,\nonumber \\
J_{D^*_s\bar{D}_s^*}(x)&=& \bar{s}(x)\gamma_\mu  c(x)\,   \bar{c}(x)\gamma^\mu  s(x) \, ,
\end{eqnarray}
\begin{eqnarray}
J_{D\bar{D}^*,+,\mu}(x)&=&\frac{1}{\sqrt{2}}\Big[\bar{u}(x)i\gamma_5c(x)\, \bar{c}(x)\gamma_\mu d(x)-\bar{u}(x)\gamma_\mu c(x)\, \bar{c}(x)i\gamma_5 d(x) \Big] \, ,\nonumber\\
J_{D\bar{D}_s^*,+,\mu}(x)&=&\frac{1}{\sqrt{2}}\Big[\bar{q}(x)i\gamma_5c(x)\, \bar{c}(x)\gamma_\mu s(x)-\bar{q}(x)\gamma_\mu c(x)\, \bar{c}(x)i\gamma_5 s(x) \Big] \, ,\nonumber\\
J_{D_s\bar{D}_s^*,+,\mu}(x)&=&\frac{1}{\sqrt{2}}\Big[\bar{s}(x)i\gamma_5c(x)\, \bar{c}(x)\gamma_\mu s(x)-\bar{s}(x)\gamma_\mu c(x)\, \bar{c}(x)i\gamma_5 s(x) \Big] \, ,
\end{eqnarray}
\begin{eqnarray}
J_{D\bar{D}^*,-,\mu}(x)&=&\frac{1}{\sqrt{2}}\Big[\bar{u}(x)i\gamma_5c(x)\, \bar{c}(x)\gamma_\mu d(x)+\bar{u}(x)\gamma_\mu c(x)\, \bar{c}(x)i\gamma_5 d(x) \Big] \, ,\nonumber\\
J_{D\bar{D}_s^*,-,\mu}(x)&=&\frac{1}{\sqrt{2}}\Big[\bar{q}(x)i\gamma_5c(x)\, \bar{c}(x)\gamma_\mu s(x)+\bar{q}(x)\gamma_\mu c(x)\, \bar{c}(x)i\gamma_5 s(x) \Big] \, ,\nonumber\\
J_{D_s\bar{D}_s^*,-,\mu}(x)&=&\frac{1}{\sqrt{2}}\Big[\bar{s}(x)i\gamma_5c(x)\, \bar{c}(x)\gamma_\mu s(x)+\bar{s}(x)\gamma_\mu c(x)\, \bar{c}(x)i\gamma_5 s(x) \Big] \, ,
\end{eqnarray}
\begin{eqnarray}
J_{D^*\bar{D}^*,-,\mu\nu}(x)&=&\frac{1}{\sqrt{2}}\Big[\bar{u}(x)\gamma_\mu c(x)\, \bar{c}(x)\gamma_\nu d(x)-\bar{u}(x)\gamma_\nu c(x)\, \bar{c}(x) \gamma_\mu d(x) \Big] \, ,\nonumber\\
J_{D^*\bar{D}_s^*,-,\mu\nu}(x)&=&\frac{1}{\sqrt{2}}\Big[\bar{q}(x)\gamma_\mu c(x)\, \bar{c}(x)\gamma_\nu s(x)-\bar{q}(x)\gamma_\nu c(x)\, \bar{c}(x) \gamma_\mu s(x) \Big] \, ,\nonumber\\
J_{D_s^*\bar{D}_s^*,-,\mu\nu}(x)&=&\frac{1}{\sqrt{2}}\Big[\bar{s}(x)\gamma_\mu c(x)\, \bar{c}(x)\gamma_\nu s(x)-\bar{s}(x)\gamma_\nu c(x)\, \bar{c}(x) \gamma_\mu s(x) \Big] \, ,
\end{eqnarray}
\begin{eqnarray}
J_{D^*\bar{D}^*,+,\mu\nu}(x)&=&\frac{1}{\sqrt{2}}\Big[\bar{u}(x)\gamma_\mu c(x)\, \bar{c}(x)\gamma_\nu d(x)+\bar{u}(x)\gamma_\nu c(x)\, \bar{c}(x) \gamma_\mu d(x) \Big] \, ,\nonumber\\
J_{D^*\bar{D}_s^*,+,\mu\nu}(x)&=&\frac{1}{\sqrt{2}}\Big[\bar{q}(x)\gamma_\mu c(x)\, \bar{c}(x)\gamma_\nu s(x)+\bar{q}(x)\gamma_\nu c(x)\, \bar{c}(x) \gamma_\mu s(x) \Big] \, ,\nonumber\\
J_{D_s^*\bar{D}_s^*,+,\mu\nu}(x)&=&\frac{1}{\sqrt{2}}\Big[\bar{s}(x)\gamma_\mu c(x)\, \bar{c}(x)\gamma_\nu s(x)+\bar{s}(x)\gamma_\nu c(x)\, \bar{c}(x) \gamma_\mu s(x) \Big] \, ,
\end{eqnarray}
and $q=u$, $d$. We construct  the color-singlet-color-singlet type local currents to interpolate the tetraquark molecular states, or more precisely, the color-singlet-color-singlet type tetraquark states \cite{WZG-Z3900-mole-10,WZG-Z4020-mole-10}.
In the isospin limit, the current operators with the  symbolic quark structures $\bar{c}c\bar{d}u$, $\bar{c}c\bar{u}d$, $\bar{c}c\frac{\bar{u}u-\bar{d}d}{\sqrt{2}}$, $\bar{c}c\frac{\bar{u}u+\bar{d}d}{\sqrt{2}}$ couple potentially  to the hidden-charm
tetraquark molecular  states with degenerated  masses, the current operators with the isospin $I=1$ and $0$ lead to the same QCD sum rules, and we will not distinguish  the isospin.

 Under parity transform $\widehat{P}$, the current  operators have the  properties,
\begin{eqnarray}
\widehat{P} J(x)\widehat{P}^{-1}&=&+J(\tilde{x}) \, , \nonumber\\
\widehat{P} J_\mu(x)\widehat{P}^{-1}&=&-J^\mu(\tilde{x}) \, , \nonumber\\
\widehat{P} J_{-,\mu\nu}(x)\widehat{P}^{-1}&=&-J_{-,}{}^{\mu\nu}(\tilde{x}) \, , \nonumber\\
\widehat{P} J_{+,\mu\nu}(x)\widehat{P}^{-1}&=&+J_{+,}{}^{\mu\nu}(\tilde{x}) \, ,
\end{eqnarray}
where  $x^\mu=(t,\vec{x})$ and $\tilde{x}^\mu=(t,-\vec{x})$.

Only the current operators or tetraquark molecular states  with the symbolic quark structures $\bar{c}c\frac{\bar{u}u-\bar{d}d}{\sqrt{2}}$, $\bar{c}c\frac{\bar{u}u+\bar{d}d}{\sqrt{2}}$ and $\bar{c}c\bar{s}s$ have definite charge conjugation. We assume  that other tetraquark molecular states have the same charge conjugation as their corresponding  charge-neutral partners.
  Under charge conjugation transform $\widehat{C}$, the currents $J(x)$, $J_\mu(x)$ and $J_{\mu\nu}(x)$ have the properties,
\begin{eqnarray}
\widehat{C}J(x)\widehat{C}^{-1}&=&+ J(x) \, , \nonumber\\
\widehat{C}J_{\pm,\mu}(x)\widehat{C}^{-1}&=&\pm J_{\pm,\mu}(x)  \, , \nonumber\\
\widehat{C}J_{\pm,\mu\nu}(x)\widehat{C}^{-1}&=&\pm J_{\pm,\mu\nu}(x)  \, .
\end{eqnarray}

\begin{table}
\begin{center}
\begin{tabular}{|c|c|c|c|c|c|c|c|c|}\hline\hline
$Z_c$($X_c$)                                                            & $J^{PC}$  & Currents              \\ \hline

$D^*\bar{D}^*$                                                          & $0^{++}$  & $J_{D^*\bar{D}^*}(x)$            \\

$D^*\bar{D}_s^*$                                                        & $0^{++}$  & $J_{D^*\bar{D}_s^*}(x)$             \\

$D_s^*\bar{D}_s^*$                                                      & $0^{++}$  & $J_{D_s^*\bar{D}_s^*}(x)$            \\ \hline

$D\bar{D}^*-D^*\bar{D}$                                                 & $1^{++}$  & $J_{D\bar{D}^*,+,\mu}(x)$           \\

$D\bar{D}_s^*-D^*\bar{D}_s$                                             & $1^{++}$  & $J_{D\bar{D}_s^*,+,\mu}(x)$              \\

$D_s\bar{D}_s^*-D_s^*\bar{D}_s$                                         & $1^{++}$  &  $J_{D_s\bar{D}_s^*,+,\mu}(x)$            \\ \hline

$D\bar{D}^*+D^*\bar{D}$                                                 & $1^{+-}$  & $J_{D\bar{D}^*,-,\mu}(x)$            \\

$D\bar{D}_s^*+D^*\bar{D}_s$                                             & $1^{+-}$  & $J_{D\bar{D}_s^*,-,\mu}(x)$                \\

$D_s\bar{D}_s^*+D_s^*\bar{D}_s$                                         & $1^{+-}$  & $J_{D_s\bar{D}_s^*,-,\mu}(x)$            \\  \hline

$D^*\bar{D}^*$                                                          & $1^{+-}$  & $J_{D^*\bar{D}^*,-,\mu\nu}(x)$   \\

$D^*\bar{D}_s^*$                                                        & $1^{+-}$  & $J_{D^*\bar{D}_s^*,-,\mu\nu}(x)$             \\

$D_s^*\bar{D}_s^*$                                                      & $1^{+-}$  & $J_{D_s^*\bar{D}_s^*,-,\mu\nu}(x)$            \\ \hline

$D^*\bar{D}^*$                                                          & $2^{++}$  & $J_{D^*\bar{D}^*,+,\mu\nu}(x)$               \\

$D^*\bar{D}_s^*$                                                        & $2^{++}$  & $J_{D^*\bar{D}_s^*,+,\mu\nu}(x)$              \\

$D_s^*\bar{D}_s^*$                                                      & $2^{++}$  & $J_{D_s^*\bar{D}_s^*,+,\mu\nu}(x)$         \\
\hline\hline
\end{tabular}
\end{center}
\caption{ The quark structures and corresponding current operators  for the hidden-charm tetraquark molecular  states. }\label{Current-Table}
\end{table}

At the hadron side, we  insert  a complete set of intermediate hadronic states with
the same quantum numbers as the current operators $J(x)$, $J_\mu(x)$ and $J_{\mu\nu}(x)$ into the
correlation functions $\Pi(p)$, $\Pi_{\mu\nu}(p)$ and $\Pi_{\mu\nu\alpha\beta}(p)$   to obtain the hadronic representation
\cite{SVZ79,Reinders85}, and isolate the ground state hidden-charm tetraquark molecule contributions,
\begin{eqnarray}
\Pi(p)&=&\frac{\lambda_{Z_+}^2}{M_{Z_+}^2-p^2} +\cdots \nonumber\\
&=&\Pi_{+}(p^2) \, ,\nonumber
\end{eqnarray}
\begin{eqnarray}
\Pi_{\mu\nu}(p)&=&\frac{\lambda_{Z_+}^2}{M_{Z_+}^2-p^2}\left( -g_{\mu\nu}+\frac{p_{\mu}p_{\nu}}{p^2}\right) +\cdots \nonumber\\
&=&\Pi_{+}(p^2)\left( -g_{\mu\nu}+\frac{p_{\mu}p_{\nu}}{p^2}\right)+\cdots \, ,\nonumber
\end{eqnarray}
\begin{eqnarray}
\Pi_{-,\mu\nu\alpha\beta}(p)&=&\frac{\lambda_{ Z_+}^2}{M_{Z_+}^2\left(M_{Z_+}^2-p^2\right)}\left(p^2g_{\mu\alpha}g_{\nu\beta} -p^2g_{\mu\beta}g_{\nu\alpha} -g_{\mu\alpha}p_{\nu}p_{\beta}-g_{\nu\beta}p_{\mu}p_{\alpha}+g_{\mu\beta}p_{\nu}p_{\alpha}+g_{\nu\alpha}p_{\mu}p_{\beta}\right) \nonumber\\
&&+\frac{\lambda_{ Z_-}^2}{M_{Z_-}^2\left(M_{Z_-}^2-p^2\right)}\left( -g_{\mu\alpha}p_{\nu}p_{\beta}-g_{\nu\beta}p_{\mu}p_{\alpha}+g_{\mu\beta}p_{\nu}p_{\alpha}+g_{\nu\alpha}p_{\mu}p_{\beta}\right) +\cdots  \nonumber\\
&=&\widetilde{\Pi}_{+}(p^2)\left(p^2g_{\mu\alpha}g_{\nu\beta} -p^2g_{\mu\beta}g_{\nu\alpha} -g_{\mu\alpha}p_{\nu}p_{\beta}-g_{\nu\beta}p_{\mu}p_{\alpha}+g_{\mu\beta}p_{\nu}p_{\alpha}+g_{\nu\alpha}p_{\mu}p_{\beta}\right) \nonumber\\
&&+\widetilde{\Pi}_{-}(p^2)\left( -g_{\mu\alpha}p_{\nu}p_{\beta}-g_{\nu\beta}p_{\mu}p_{\alpha}+g_{\mu\beta}p_{\nu}p_{\alpha}+g_{\nu\alpha}p_{\mu}p_{\beta}\right) \, ,\nonumber
\end{eqnarray}
\begin{eqnarray}
\Pi_{+,\mu\nu\alpha\beta}(p)&=&\frac{\lambda_{ Z_+}^2}{M_{Z_+}^2-p^2}\left( \frac{\widetilde{g}_{\mu\alpha}\widetilde{g}_{\nu\beta}+\widetilde{g}_{\mu\beta}\widetilde{g}_{\nu\alpha}}{2}-\frac{\widetilde{g}_{\mu\nu}\widetilde{g}_{\alpha\beta}}{3}\right) +\cdots \, \, , \nonumber \\
&=&\Pi_{+}(p^2)\left( \frac{\widetilde{g}_{\mu\alpha}\widetilde{g}_{\nu\beta}+\widetilde{g}_{\mu\beta}\widetilde{g}_{\nu\alpha}}{2}-\frac{\widetilde{g}_{\mu\nu}\widetilde{g}_{\alpha\beta}}{3}\right) +\cdots\, ,
\end{eqnarray}
where $\widetilde{g}_{\mu\nu}=g_{\mu\nu}-\frac{p_{\mu}p_{\nu}}{p^2}$, and the $Z$ represents the tetraquark molecular states $Z_c$, $X_c$, $Z_{cs}$, etc.  We add the subscripts $\pm$  in the hidden-charm tetraquark molecular states $Z_{\pm}$ and the components of the correlation functions $\Pi_{\pm}(p^2)$ and $\widetilde{\Pi}_{\pm}(p^2)$ to represent  the positive  and negative parity contributions, respectively.
The   pole residues   $\lambda_{Z_\pm}$ are defined by
\begin{eqnarray}
 \langle 0|J(0)|Z_+(p)\rangle &=&\lambda_{Z_+}\, , \nonumber\\
 \langle 0|J_\mu(0)|Z_+(p)\rangle &=&\lambda_{Z_+}\varepsilon_\mu\, , \nonumber\\
  \langle 0|J_{-,\mu\nu}(0)|Z_+(p)\rangle &=& \frac{\lambda_{Z_+}}{M_{Z_+}} \, \varepsilon_{\mu\nu\alpha\beta} \, \varepsilon^{\alpha}p^{\beta}\, , \nonumber\\
 \langle 0|J_{-,\mu\nu}(0)|Z_-(p)\rangle &=&\frac{\lambda_{Z_-}}{M_{Z_-}} \left(\varepsilon_{\mu}p_{\nu}-\varepsilon_{\nu}p_{\mu} \right)\, , \nonumber\\
  \langle 0|J_{+,\mu\nu}(0)|Z_+(p)\rangle &=& \lambda_{Z_+}\, \varepsilon_{\mu\nu} \, ,
\end{eqnarray}
where the  $\varepsilon_{\mu/\alpha}$ and $\varepsilon_{\mu\nu}$ are the polarization vectors of the molecular states.
In this article, we choose the components $\Pi_{+}(p^2)$ and $p^2\widetilde{\Pi}_{+}(p^2)$ to study the scalar, axialvector and tensor hidden-charm tetraquark molecular states with the QCD sum rules.   In Table \ref{Current-Table}, we present the inner quark structures and corresponding interpolating current operators for the hidden-charm tetraquark molecular states.

We maybe worry about the contaminations from the two-meson scattering states, as the quantum field theory does not forbid the couplings tween the color-singlet-color-singlet type four-quark currents and the two-meson scattering states if they have the same quantum numbers. At the hadron side of the QCD sum rules, we can take account of the contributions of  both the tetraquark molecular states and two-meson scattering states tentatively.   In Ref.\cite{WangZG-Landau}, we  take  an axialvector current and a tensor current as an example besides presenting detailed discussions  to illustrate that the two-meson scattering states alone cannot saturate the QCD sum rules,  while the tetraquark molecular states alone  can saturate the QCD sum rules,  the net effects of the two-meson scattering states amount to give a finite width to the tetraquark molecular states, which can be absorbed into the pole residues safely without affecting the predicted  molecule masses.

At the QCD side of the correlation functions $\Pi(p)$, $\Pi_{\mu\nu}(p)$ and $\Pi_{\mu\nu\alpha\beta}(p)$, we contract the quark fields with the Wick's theorem and observe that there are  two  heavy  quark propagators and two light quark propagators.  If each heavy quark line emits a gluon and each light quark line contributes  a quark-antiquark  pair, we obtain a quark-gluon  operator $GG\bar{q}q \bar{q}q$   of dimension 10, so we should calculate  the vacuum condensates at least
up to dimension 10 to judge the convergent behavior of the operator product expansion.

In the present work, we carry out the operator product expansion  up to the vacuum condensates of dimension $10$ consistently, and take  account of the vacuum condensates $\langle\bar{q}q\rangle$, $\langle\frac{\alpha_{s}GG}{\pi}\rangle$, $\langle\bar{q}g_{s}\sigma Gq\rangle$, $\langle\bar{q}q\rangle^2$, $g_s^2\langle\bar{q}q\rangle^2$,
$\langle\bar{q}q\rangle \langle\frac{\alpha_{s}GG}{\pi}\rangle$,  $\langle\bar{q}q\rangle  \langle\bar{q}g_{s}\sigma Gq\rangle$,
$\langle\bar{q}g_{s}\sigma Gq\rangle^2$ and $\langle\bar{q}q\rangle^2 \langle\frac{\alpha_{s}GG}{\pi}\rangle$,  where we choose the $q$ to represent the $u$, $d$ or $s$ quark together for simplicity. For the light flavor $SU(3)$ breaking effects, we neglect the small masses of the $u$ and $d$ quarks, and take account of  the terms proportional to the mass of the $s$-quark, in other words, the terms which are linear in the strange quark mass $m_s$.

 There are terms  $\langle\bar{q}_j\sigma_{\mu\nu}q_i \rangle$ and $\langle\bar{q}_j\gamma_{\mu}q_i\rangle$ in the full light quark propagators \cite{WangHuangtao-2014-PRD}, which  absorb the gluons  emitted from the other quark lines to form $\langle\bar{q}_j g_s G_{\alpha\beta} \sigma_{\mu\nu} q_i \rangle$ and $\langle\bar{q}_j\gamma_{\mu}q_ig_s D_\nu G_{\alpha\beta}\rangle$, and  contribute  to the mixed condensates and four-quark condensates $\langle\bar{q}g_s\sigma G q\rangle$ and $g_s^2\langle\bar{q}q\rangle^2$, respectively, where $D_\alpha=\partial_\alpha-ig_sG^n_\alpha t^n$, $t^n=\frac{\lambda^n}{2}$.
 The four-quark condensates $g_s^2\langle \bar{q}q\rangle^2$ come from the terms
$\langle \bar{q}\gamma_\mu t^a q g_s D_\eta G^a_{\lambda\tau}\rangle$, $\langle\bar{q}_jD^{\dagger}_{\mu}D^{\dagger}_{\nu}D^{\dagger}_{\alpha}q_i\rangle$  and
$\langle\bar{q}_jD_{\mu}D_{\nu}D_{\alpha}q_i\rangle$,  the strong fine-structure  constant $\alpha_s(\mu)=\frac{g_s^2(\mu)}{4\pi}$ appears  at the tree level, and cannot be absorbed into the vacuum condensates, although the contributions of the $g_s^2\langle\bar{q}q\rangle^2$ are tiny.

In calculations, we can see clearly that  all the Feynman diagrams can be divided into factorizable diagrams and nonfactorizable diagrams  in the color space.  Lucha, Melikhov and Sazdjian assert that the contributions at  the order $\mathcal{O}(\alpha_s^k)$ with $k\leq1$ in the operator product expansion, which can be factorized into two color-neutral clusters, are exactly canceled out by the two-meson scattering states, while the tetraquark (molecular) states begin receive contributions at the order $\mathcal{O}(\alpha_s^2)$ \cite{Chu-Sheng-3}. In fact, the nonfactorizable Feynman diagrams begin to appear at the order $\mathcal{O}(\alpha_s^1)$ rather than at the order $\mathcal{O}(\alpha_s^2)$ \cite{WangZG-Landau}.

It is useless to distinguish the
factorizable and nonfactorizable properties of the Feynman diagrams, we can only obtain information about the
short-distance (perturbative) and long-distance (nonperturbative) contributions.   In the operator product expansion,
the short-distance (perturbative) contributions above a certain
energy scale $\mu$ are included in the Wilson's coefficients, the long-distance (nonperturbative) contributions below this
special energy scale $\mu$ are included in the vacuum condensates. In general, we can choose any
energy scales at which the perturbative QCD calculations are feasible.
All  the Feynman diagrams make contributions to the  tetraquark molecular states, in other words,  the   tetraquark molecular states begin to receive contributions at the order $\mathcal{O}(\alpha_s^0/\alpha_s^1)$ rather than at the order $\mathcal{O}(\alpha_s^2)$ \cite{WangZG-Landau}.

As we have obtained both the hadron spectral representations and the QCD spectral representations, now we match  the hadron side with the QCD  side of the components $\Pi_{+}(p^2)$ and $p^2\widetilde{\Pi}_{+}(p^2)$ below the continuum thresholds   $s_0$ and perform Borel transform  in regard  to
the variable  $P^2=-p^2$ to obtain  the  QCD sum rules:
\begin{eqnarray}\label{QCDSR}
\lambda^2_{Z_+}\, \exp\left(-\frac{M^2_{Z_+}}{T^2}\right)= \int_{4m_c^2}^{s_0} ds\, \rho_{QCD}(s) \, \exp\left(-\frac{s}{T^2}\right) \, ,
\end{eqnarray}
 the lengthy QCD spectral densities $\rho_{QCD}(s)$ are neglected for simplicity. The interested readers who want to obtain the analytical expressions of the QCD
 spectral densities can contact with me via E-mail directly.

We derive Eq.\eqref{QCDSR} in regard  to  $\tau=\frac{1}{T^2}$,  and obtain the QCD sum rules for
 the masses of the  scalar,  axialvector  and tensor hidden-charm tetraquark molecular states $Z_+$ without strange, with strange and with hidden-strange,
 \begin{eqnarray}
 M^2_{Z_+}&=& -\frac{\int_{4m_c^2}^{s_0} ds\frac{d}{d \tau}\rho(s)\exp\left(-\tau s \right)}{\int_{4m_c^2}^{s_0} ds \rho(s)\exp\left(-\tau s\right)}\mid_{\tau=\frac{1}{T^2}}\, .
\end{eqnarray}

\section{Numerical results and discussions}
Now let us choose the input parameters at the QCD side. We take  the standard values of the  vacuum condensates
$\langle\bar{q}q \rangle=-(0.24\pm 0.01\, \rm{GeV})^3$,  $\langle\bar{s}s \rangle=(0.8\pm0.1)\langle\bar{q}q \rangle$,
 $\langle\bar{q}g_s\sigma G q \rangle=m_0^2\langle \bar{q}q \rangle$, $\langle\bar{s}g_s\sigma G s \rangle=m_0^2\langle \bar{s}s \rangle$,
$m_0^2=(0.8 \pm 0.1)\,\rm{GeV}^2$, $\langle \frac{\alpha_s
GG}{\pi}\rangle=(0.33\,\rm{GeV})^4$    at the energy scale  $\mu=1\, \rm{GeV}$
\cite{SVZ79,Reinders85,Colangelo-Review}, and  take the $\overline{MS}$ masses $m_{c}(m_c)=(1.275\pm0.025)\,\rm{GeV}$
 and $m_s(\mu=2\,\rm{GeV})=(0.095\pm0.005)\,\rm{GeV}$
 from the Particle Data Group \cite{PDG}.
Furthermore,  we take account of
the energy-scale dependence of  the quark condensates, mixed quark condensates and $\overline{MS}$ masses in regard  to  the renormalization group equation \cite{Narison-mix},
 \begin{eqnarray}
 \langle\bar{q}q \rangle(\mu)&=&\langle\bar{q}q\rangle({\rm 1 GeV})\left[\frac{\alpha_{s}({\rm 1 GeV})}{\alpha_{s}(\mu)}\right]^{\frac{12}{33-2n_f}}\, , \nonumber\\
 \langle\bar{s}s \rangle(\mu)&=&\langle\bar{s}s \rangle({\rm 1 GeV})\left[\frac{\alpha_{s}({\rm 1 GeV})}{\alpha_{s}(\mu)}\right]^{\frac{12}{33-2n_f}}\, , \nonumber\\
 \langle\bar{q}g_s \sigma Gq \rangle(\mu)&=&\langle\bar{q}g_s \sigma Gq \rangle({\rm 1 GeV})\left[\frac{\alpha_{s}({\rm 1 GeV})}{\alpha_{s}(\mu)}\right]^{\frac{2}{33-2n_f}}\, ,\nonumber\\
  \langle\bar{s}g_s \sigma Gs \rangle(\mu)&=&\langle\bar{s}g_s \sigma Gs \rangle({\rm 1 GeV})\left[\frac{\alpha_{s}({\rm 1 GeV})}{\alpha_{s}(\mu)}\right]^{\frac{2}{33-2n_f}}\, ,\nonumber\\
m_c(\mu)&=&m_c(m_c)\left[\frac{\alpha_{s}(\mu)}{\alpha_{s}(m_c)}\right]^{\frac{12}{33-2n_f}} \, ,\nonumber\\
m_s(\mu)&=&m_s({\rm 2GeV} )\left[\frac{\alpha_{s}(\mu)}{\alpha_{s}({\rm 2GeV})}\right]^{\frac{12}{33-2n_f}}\, ,\nonumber\\
\alpha_s(\mu)&=&\frac{1}{b_0t}\left[1-\frac{b_1}{b_0^2}\frac{\log t}{t} +\frac{b_1^2(\log^2{t}-\log{t}-1)+b_0b_2}{b_0^4t^2}\right]\, ,
\end{eqnarray}
  where $t=\log \frac{\mu^2}{\Lambda^2}$, $b_0=\frac{33-2n_f}{12\pi}$, $b_1=\frac{153-19n_f}{24\pi^2}$, $b_2=\frac{2857-\frac{5033}{9}n_f+\frac{325}{27}n_f^2}{128\pi^3}$,  $\Lambda=213\,\rm{MeV}$, $296\,\rm{MeV}$  and  $339\,\rm{MeV}$ for the quark flavors  $n_f=5$, $4$ and $3$, respectively  \cite{PDG}.
As we investigate  the hidden-charm tetraquark molecular states without strange, with strange and with hidden-strange, it is better to  choose the quark flavor numbers $n_f=4$, and evolve the QCD spectral densities $\rho_{QCD}(s)$ to the ideal energy scales $\mu$ to extract the molecule masses.

In the heavy quark limit, the hidden-bottom or hidden-charm tetraquark (molecular) states (or the exotic $X$, $Y$, $Z$ states) can be described by  a double-well potential model,  the heavy  quark $Q$ (heavy antiquark $\overline{Q}$) serves as a static well potential, then we introduce the effective heavy quark masses $\mathbb{M}_Q$ and divide the tetraquark (molecular) states into both the heavy degrees of freedom and light degrees of freedom. If we neglect the small $u$ and $d$ quark masses, we obtain the heavy degrees of freedom
$2{\mathbb{M}}_Q$ and light degrees of freedom $\mu=\sqrt{M^2_{X/Y/Z}-(2{\mathbb{M}}_Q)^2}$.  We can also introduce the effective $s$-quark mass $\mathbb{M}_s$ to take  account of  the light flavor $SU(3)$ breaking effects.

 Analysis of the $J/\psi$ and $\Upsilon$ mass spectrum
with the famous   Cornell potential or Coulomb-potential-plus-linear-potential leads to the constituent quark masses $m_c=1.84\,\rm{GeV}$ and $m_b=5.17\,\rm{GeV}$ \cite{Cornell}, we can set the effective $c$-quark mass  to be  the constituent quark mass ${\mathbb{M}}_c=m_c=1.84\,\rm{GeV}$.   The old value ${\mathbb{M}}_c=1.84\,\rm{GeV}$ and updated value  ${\mathbb{M}}_c=1.85\,\rm{GeV}$, which are fitted in the QCD sum rules for the hidden-charm tetraquark molecular states,  are all consistent with the  constituent quark mass $m_c=1.84\,\rm{GeV}$ \cite{WZG-Z3900-mole-10,WZG-Z4020-mole-10, WZG-Y4260-mol-10-CPC}. We can choose the value ${\mathbb{M}}_c=1.84\,\rm{GeV}$ to determine the ideal energy scales of the QCD spectral densities, and add an uncertainty $\delta\mu=\pm0.1\,\rm{GeV}$ to account for the difference between the values ${\mathbb{M}}_c=1.84\,\rm{GeV}$ and $1.85\,\rm{GeV}$. Considering for  the light flavor $SU(3)$ mass breaking effects, we take the modified energy scale formula $\mu=\sqrt{M^2_{X/Y/Z}-(2{\mathbb{M}}_c)^2}-k\,\mathbb{M}_s$, with $k=0$, $1$, $2$ and $\mathbb{M}_s=0.2\,\rm{GeV}$, which is  proved work well.

The continuum threshold parameters should be large enough to take  account of the contributions of the ground state tetraquark molecular states fully but small enough to avoid the contaminations from the higher resonances  and continuum states.
 For the conventional heavy mesons and quarkonia, the energy gaps between the ground states and the first radial excited states are about $0.5\sim 0.6\,\rm{GeV}$ \cite{PDG}, we tentatively choose the continuum threshold parameters as $\sqrt{s_0}=M_{Z}+0.4\sim 0.7\,\rm{GeV}$ and vary the continuum threshold parameters $s_0$ and Borel parameters $T^2$ to satisfy
the  four   criteria:\\
$\bullet$ Pole dominance at the hadron  side;\\
$\bullet$ Convergence of the operator product expansion at the QCD side;\\
$\bullet$ Appearance of the Borel platforms;\\
$\bullet$ Satisfying the  energy scale formula,\\
  via trial  and error.
The first two criteria are two basic  criteria for the QCD sum rules, we should satisfy them to obtain reliable QCD sum rules.
In addition, we should obtain Borel platforms to avoid additional uncertainties from the Borel parameters. The energy scale formula is a double-edged sword, although it can enhance the pole contributions remarkably at the hadron side and improve the convergent behaviors of the operator product expansion remarkably
 at the QCD side,   it also  puts forward  a strict constraint to obey for  the energy scales of the QCD spectral densities.

Finally, we obtain the Borel windows, continuum threshold parameters, energy scales of the QCD spectral densities,  pole contributions, and the contributions of the vacuum condensates of dimension $10$, the highest dimension, which are shown explicitly in Table \ref{BorelP}.
From the Table,  we can see clearly  that the pole contributions are about $(40-60)\%$ at the hadron side, while the central values are larger than $50\%$, the pole dominance condition  is well satisfied.
At the QCD side, the contributions of the vacuum condensates of dimension $10$ are $|D(10)|\leq 1 \%$ or $\ll 1\%$, the convergent behaviors  of the operator product  expansion are  very good.

We take  account of  all the uncertainties of the input parameters including the uncertainty of the energy scale of the QCD spectral densities $\delta\mu=0.1\,\rm{GeV}$,  and obtain the masses and pole residues of the scalar, axialvector, tensor hidden-charm  tetraquark molecular  states without strange, with strange and with hidden-strange, which are shown explicitly in Table \ref{mass-Table}. From  Tables \ref{BorelP}--\ref{mass-Table}, we can see clearly that the modified energy scale formula $\mu=\sqrt{M^2_{X/Y/Z}-(2{\mathbb{M}}_c)^2}-k\,\mathbb{M}_s$ is well satisfied.

\begin{figure}
\centering
\includegraphics[totalheight=6cm,width=7cm]{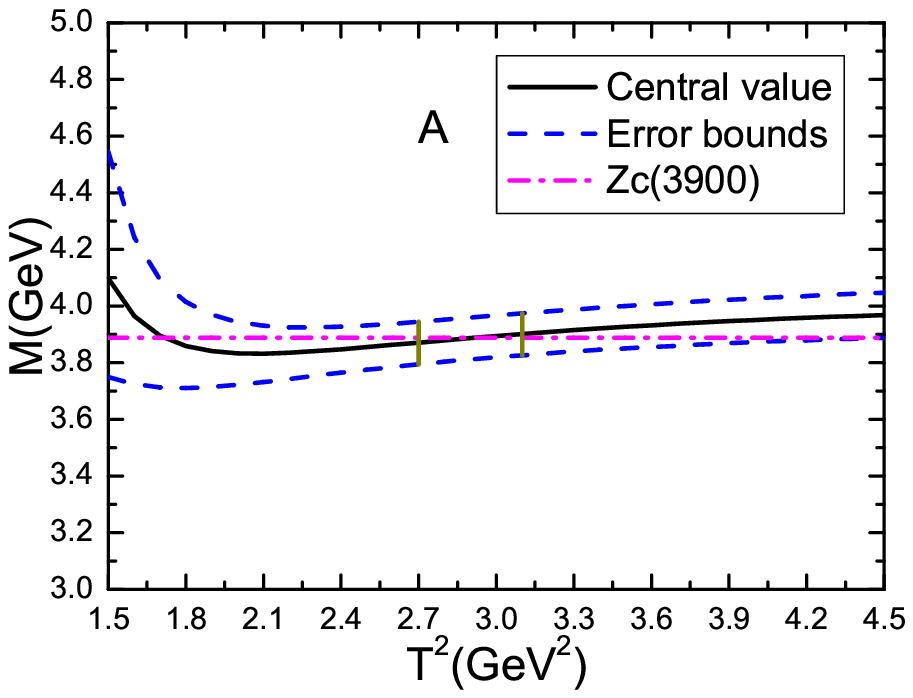}
\includegraphics[totalheight=6cm,width=7cm]{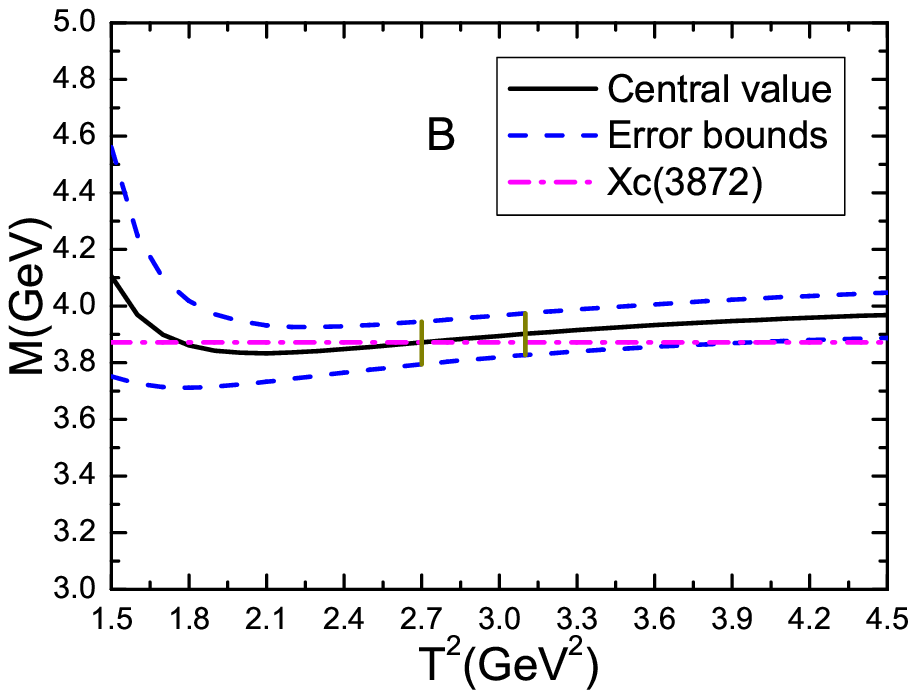}
\includegraphics[totalheight=6cm,width=7cm]{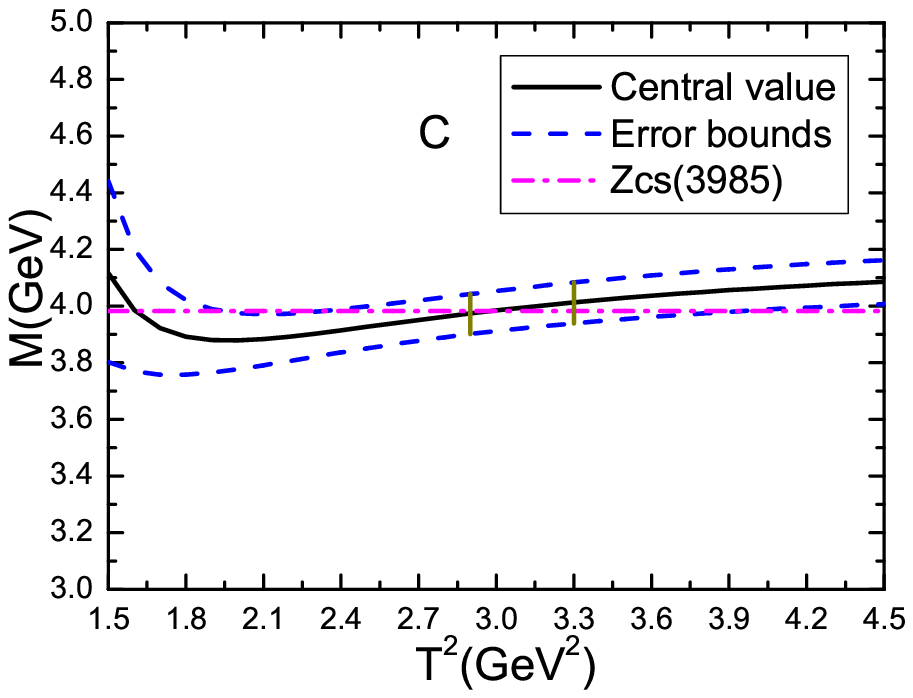}
\includegraphics[totalheight=6cm,width=7cm]{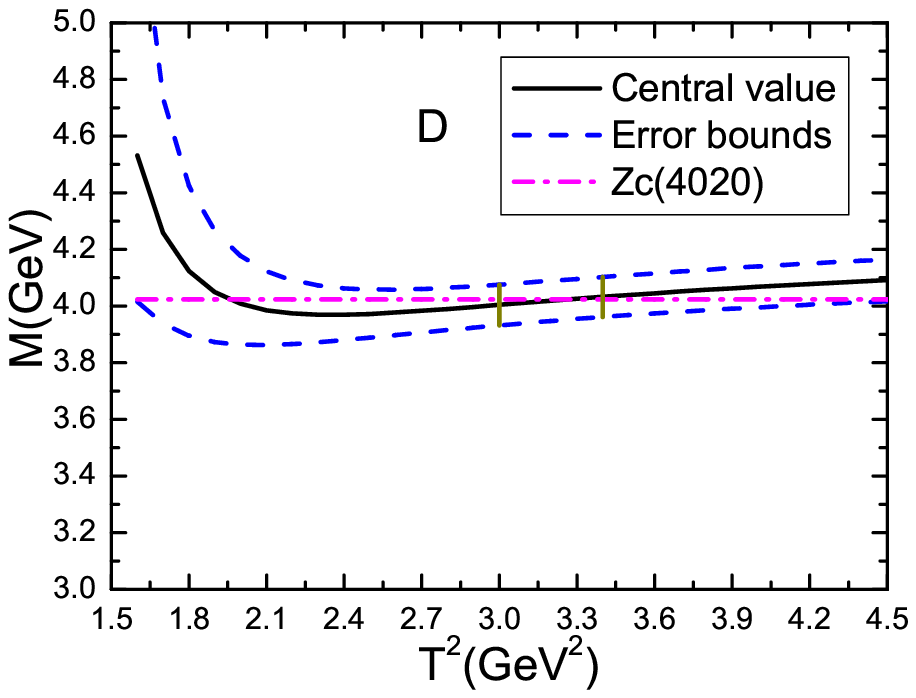}
  \caption{ The masses of the tetraquark  molecular states with variations of the Borel parameters $T^2$, where the $A$, $B$, $C$  and $D$ correspond  to the axialvector tetraquark molecular states $D\bar{D}^*+D^*\bar{D}$, $D\bar{D}^*-D^*\bar{D}$, $D\bar{D}_s^*+D^*\bar{D}_s$   and $D^*\bar{D}^*$, respectively.    }\label{mass-Zc-Zcs}
\end{figure}

 In  Fig.\ref{mass-Zc-Zcs}, we plot the masses of the  axialvector tetraquark molecular states $D\bar{D}^*+D^*\bar{D}$, $D\bar{D}^*-D^*\bar{D}$, $D\bar{D}_s^*+D^*\bar{D}_s$   and $D^*\bar{D}^*$ with variations of the Borel parameters at much larger ranges than the Borel widows as an example. From the figure, we can see plainly that there appear very flat platforms in the Borel windows indeed, where the regions between the two short vertical lines are the Borel windows.       Now the four criteria of the QCD sum rules are all satisfied, and we expect to make reliable predictions.

In Fig.\ref{mass-Zc-Zcs}, we also present the experimental values of the masses of the $Z_c(3900)$, $X_c(3872)$, $Z_{cs}(3985)$ and $Z_c(4020)$ \cite{PDG,BES3985}, from the figure, we can see explicitly that the predicted masses are in excellent agreement   with the experimental data. The present calculations support assigning the $Z_c(3900)$, $X_c(3872)$, $Z_{cs}(3985)$ and $Z_c(4020)$ to be the $D\bar{D}^*+D^*\bar{D}$, $D\bar{D}^*-D^*\bar{D}$, $D\bar{D}_s^*+D^*\bar{D}_s$   and $D^*\bar{D}^*$ tetraquark molecular states with the quantum numbers $J^{PC}=1^{+-}$, $1^{++}$, $1^{+-}$ and $1^{+-}$, respectively.
In Table \ref{Assignments-Table}, we present the possible assignments of the ground state hidden-charm tetraquark molecular states.

 We can reproduce the experimental values of the masses of the  $X_c(3872)$, $Z_c(3900)$, $Z_{cs}(3985)$ and $Z_c(4020)$ both in the  scenarios  of tetraquark  states \cite{Narison-3872,Azizi-Zcs3985-tetra-6,CFQiao-Zcs3985-tetra-mol-8,WangHuangtao-2014-PRD,Wang-tetra-formula,
WZG-hidden-cc-tetra-10,WZG-Z3900-two-particle-10,WZG-Zcs3985-tetra-10}  and tetraquark molecular states \cite{CFQiao-Zcs3985-tetra-mol-8,Nielsen-X3872-mol-cc-8,HXChen-Zcs3985-mol-8,WZG-Z3900-mole-10,WZG-Z4020-mole-10}. However, in the scenario of the tetraquark states, we can  accommodate much more exotic $X$ and $Z$ states, see Tables \ref{Assignments-Table-tetraquark}-\ref{Assignments-Table-tetraquark-Zcs}.
Even in the scenario of tetraquark  states, there are no rooms to accommodate the $X(3940)$, $X(4160)$, $Z_c(4100)$ and $Z_c(4200)$ without resorting to fine tuning. The $X(3940)$ and $X(4160)$ may be  the conventional $\eta_c(\rm3S)$ and $\eta_c(\rm 4S)$ states  with the quantum numbers  $J^{PC}=0^{-+}$, respectively. The $Z_c(4100)$ may be a mixing scalar  tetraquark state with the quantum numbers $J^{PC}=0^{++}$, and the $Z_c(4200)$ may be  an axialvector color-octet-color-octet  type tetraquark state with the quantum numbers $J^{PC}=1^{+-}$. For detailed discussions about this subject, one can consult Ref.\cite{WZG-hidden-cc-tetra-10}.
The Table \ref{Assignments-Table-tetraquark} and Table \ref{Assignments-Table-tetraquark-Zcs}  are taken from Ref.\cite{WZG-hidden-cc-tetra-10} and Ref.\cite{WZG-Zcs3985-tetra-10}, respectively.

 In Ref.\cite{WZG-hidden-cc-tetra-10},  we take the pseudoscalar, scalar, axialvector, vector, tensor charmed (anti)diquark operators as the fundamental  building blocks, and construct  the scalar, axialvector and tensor hidden-charm four-quark currents to explore the  mass spectrum of the ground state hidden-charm tetraquark states  with the QCD sum rules in a comprehensive way, and make possible assignments of the existing exotic $X$, $Y$, $Z$ states in the scenario of the tetraquark states.
In Ref.\cite{WZG-Z4600-tetra-10}, we construct the axialvector and tensor four-quark currents to investigate the ground state and first radially excited tetraquark states  with the quantum numbers $J^{PC}=1^{+-}$ via the QCD sum rules, and observe that we can assign the $Z_c(3900)$ and $Z_c(4430)$ as the ground state and first radially excited state of the $[uc]_S[\bar{d}\bar{c}]_A-[uc]_A[\bar{d}\bar{c}]_S$ type axialvector tetraquark states, respectively;
and assign the $Z_c(4020)$ and $Z_c(4600)$ as the ground state and  first radially  excited  $[uc]_{\tilde{A}}[\bar{d}\bar{c}]_A-[uc]_A[\bar{d}\bar{c}]_{\tilde{A}}$ type or $[uc]_A[\bar{d}\bar{c}]_A$ type axialvector tetraquark states, respectively, which are also shown in Table \ref{Assignments-Table-tetraquark}, see the last column.

In Ref.\cite{WZG-Zcs3985-tetra-10},  we  investigate  the diquark-antidiquark type axialvector tetraquark states $c\bar{c}u\bar{s}$ with the quantum numbers $J^{PC}=1^{+-}$ in the framework of  the QCD sum rules, and obtain the prediction $M_{Z_{cs}}=3.99\pm0.09\,\rm{GeV}$, which is in very good agreement with the experimental data  $3985.2^{+2.1}_{-2.0}\pm1.7\,\rm{MeV}$ from the BESIII collaboration \cite{BES3985}, and  supports assigning  the $Z_{cs}(3985)$   as the $SU(3)$ cousin of the $Z_c(3900)$ in  the scenario of tetraquark  states. Then we take account of the light flavor $SU(3)$ mass-breaking effects  to estimate the mass spectrum of the diquark-antidiquark type hidden-charm  tetraquark states with strangeness, which are shown clearly in Table \ref{Assignments-Table-tetraquark-Zcs}. As the $Z_{cs}(3985)$ has no definite charge conjugation, the assignment of the quantum numbers $J^{PC}=1^{++}$ cannot be excluded.

In Ref.\cite{WZG-Y4660-decay}, we  investigate  the two-body strong decays of the $Y(4660)$ with the QCD sum rules based on solid (or rigorous) quark-hadron quality. The predicted total decay width  is in excellent agreement with the experimental data  from the Belle collaboration \cite{Belle-Y4660-2014}, which supports assigning the $Y(4660)$ to be a vector  tetraquark state with the quantum numbers $J^{PC}=1^{--}$.
The predicted  hadronic coupling constants $ |G_{Y\psi^\prime f_0}|\gg |G_{Y J/\psi f_0}|$, which is consistent with the observation of the $Y(4660)$ in the $\psi^\prime\pi^+\pi^-$ invariant mass spectrum, and favors the $\psi^{\prime}f_0(980)$ molecule assignment \cite{WZG-Y4660-mole-10}.  Those exotic  $X$, $Y$ and $Z$ states maybe have a diquark-antidiquark type tetraquark core with the typical size of the $q\bar{q}$-type  mesons, the strong couplings  to the meson-meson pairs   lead to some tetraquark molecule Fock  components, and they maybe spend  a rather large time as the  tetraquark   molecular states, just like the $f_0(980)$ and $a_0(980)$ \cite{ReviewAmsler2}.

Generally speaking,  a hadron has definite quantum numbers and several Fock states, any current with the same quantum numbers and quark structures as a Fock state in a hadron couples potentially to this  hadron. In this respect, we can construct several currents to interpolate a hadron, or construct a current to interpolate several hadrons. We call a hadron as a tetraquark (molecular) state if its main Fock component is of the diquark-antidiquark type (color-singlet-color-singlet type),  and choose the suitable four-quark  currents to interpolate it. One possibility  is that those $X_c$ and $Z_c$ states have competitive or evenly matched  color-antitriplet-color-triplet type and  color-singlet-color-singlet type Fock  components, we can choose either  the $ \bar{\bf3}_c{\bf 3}_c$-type or  ${\bf 1}_c{\bf 1}_c$-type four-quark currents  to interpolate them.

The $Z_c(3900)$ and $Z_c(3885)$ have almost degenerated masses but quite different decay widths  from the BESIII collaboration \cite{BES3900,BES-3885}.
They are taken as the same particle  and are named as $Z_c(3900)$ by the Particle Data Group \cite{PDG}, however, without introducing the molecule Fock component, it is difficult to take  account of  the large  ratio $R_{exp} =\frac{\Gamma(Z_c(3885)\to D\bar{D}^*)}{\Gamma(Z_c(3900)\to J/\psi \pi)}=6.2 \pm 1.1 \pm 2.7 $
from  the BESIII collaboration \cite{BES-3885}. If we assign the $Z_c(3900)$ to be the  diquark-antidiquark type axialvector tetraquark state, and assign the $Z_c(3885)$
to be  the $D\bar{D}+D^{*}\bar{D}$ tetraquark molecular state, it is easy to account for the large ratio $R_{exp}$.
From Tables \ref{Assignments-Table}-\ref{Assignments-Table-tetraquark}, we can see that the $Z_c(3900)$ and $Z_c(4430)$ can be assigned to be the ground state and first radial excited axialvector tetraquark (molecular) states with the quantum numbers $J^{PC}=1^{+-}$, respectively according to the energy gap $0.59\,\rm{GeV}$, irrespective of the  scenarios  of tetraquark  states and tetraquark molecular states. There is another outcome or possibility,
there maybe exist double exotic states with the almost  degenerated masses besides the same quantum numbers, just like the $Z_c(3900)$ and $Z_c(3885)$.

In 2020, the BESIII collaboration observed an excess of events over the known contributions of the conventional charmed mesons near the $D_s^- D^{*0}$ and $D^{*-}_s D^0$ thresholds in the $K^{+}$ recoil-mass spectrum with the significance of about  $5.3\sigma$ in the processes of $e^+e^-\to K^+ (D_s^- D^{*0}+ D^{*-}_s D^0)$ \cite{BES3985}. Assuming the light flavor $SU(3)$ symmetry, we would expect that there  exist the strange partners of the $Z_c$ states,
denoted as $Z_{cs}$, with the valence quarks $c\bar{c}q\bar{s}$, therefore the new structure was named as $Z_{cs}(3985)$.
Its Breit-Wigner  mass and width are  $3985.2^{+2.1}_{-2.0}\pm1.7\,\rm{MeV}$ and $13.8^{+8.1}_{-5.2}\pm4.9\,\rm{MeV}$ respectively with an assignment of the spin-parity $J^P=1^+$  \cite{BES3985}.  At the energy about $4.1\,\rm{GeV}$ in the $K^{+}$ recoil-mass spectrum, there is also an  indiction of a broader enhancement \cite{BES3985}, which maybe  due to the $D^*\bar{D}_s^*$ molecular state, $Z_{cs}(4100)$, with the quantum numbers $J^{PC}=1^{+-}$ predicted in the present work, this  possible assignment is also presented in Table \ref{Assignments-Table}.

After the present manuscript was finished and submitted to https://arxiv.org/, the LHCb collaboration reported its new observations.
The LHCb collaboration observed two new  exotic states with the valence quarks  $c\bar{c}u\bar{s}$  decaying to the $J/\psi K^+$  final state in the process  $B^+ \to J/\psi K^+$ \cite{LHCb-Zcs4000}.  The most significant state, $Z_{cs}(4000)$, has a mass of $4003 \pm 6 {}^{+4}_{-14}\,\rm{MeV}$, a width of $131 \pm 15 \pm 26\,\rm{MeV}$, and the spin-parity
$J^P =1^+$, while the broader state, $Z_{cs}(4220)$, has a mass of $4216 \pm 24{}^{ +43}_{-30}\,\rm{MeV}$, a width of $233 \pm 52 {}^{+97}_{-73}\,\rm{MeV}$, and the spin-parity $J^P=1^+$ or $1^-$, with a $2\sigma$  difference in favor of the first hypothesis \cite{LHCb-Zcs4000}. Furthermore, the LHCb collaboration confirmed the
 four previously reported $J/\psi \phi$  states $X(4140)$, $X(4274)$, $X(4500)$, $X(4700)$, and observed two new $X$ states with the valence quarks $c\bar{c}s\bar{s}$, the $X(4685)$ and $X(4630)$ with the spin-parity $J^P=1^+$ and $1^-$, respectively \cite{LHCb-Zcs4000}.

According to the measured masses, the $Z_{cs}(3885)$ and $Z_{cs}(4000)$ can be assigned to be the same particle. However, the width of the $Z_{cs}(4000)$ is about ten times as large as the width of the $Z_{cs}(3985)$, in this respect, they are two quite different particles.   In the LHCb data of the $J/\psi K^+$ mass distribution, there is a hint of a dip at the energy about $4.1\,\rm{GeV}$ \cite{LHCb-Zcs4000}, which maybe due to the missing $Z_{cs}(4100)$,
 as there is  a  weak enhancement,  the $Z_{cs}(4100)$,  in the $K^+$ recoil-mass spectrum in the BESIII data \cite{BES3985}.
More experimental data are still needed to obtain a precise resolution, which maybe lead to a narrower width  for the $Z_{cs}(4000)$. Even if the  $Z_{cs}(3885)$ and $Z_{cs}(4000)$ are two particles, we have enough rooms to accommodate them, there maybe exist a diquark-antidiquark type and a color-singlet-color-singlet type tetraquark states, which have the masses $3.99\pm0.09\,\rm{GeV}$ \cite{WZG-Zcs3985-tetra-10} and $3.99\pm0.09\,\rm{GeV}$, respectively.

Naively, we expect that the color-singlet-color-singlet type tetraquark states have narrower widths  than the diquark-antidiquark type tetraquark states with the same valence quarks, under the condition that the corresponding mesons with the same quantum numbers as the color-neutral (or color-singlet) clusters  have narrow widths. In fact, such naive expectations only survive for the loosely bound molecular states which have large average spatial sizes, $\sqrt{\langle r^2\rangle}\sim \sqrt{\langle r^2\rangle_M}+\sqrt{\langle r^2\rangle_{M^\prime}}$, where the $M$ and $M^\prime$ denote the conventional mesons. In the QCD sum rules, we choose the local currents, the tetraquark molecular states are also compact objects and have the average spatial sizes as the tetraquark states, their widths are not necessary to be narrower than that of the tetraquark states. It is necessary to  investigate their decay widths via the QCD sum rules directly  to make  more reliable assignments.

There is no room to accommodate the $Z_{cs}(4220)$ as the $J^P=1^+$ exotic state  both in the    scenarios of tetraquark  state and tetraquark molecular state, see Table \ref{Assignments-Table} and Table \ref{Assignments-Table-tetraquark-Zcs}. On the other hand, if we assign the $Z_{cs}(4220)$ to be  the diquark-antidiquark type  tetraquark state or color-singlet-color-singlet type  molecular state with the spin-parity $J^{P}=1^-$, it is also difficult to reproduce the experimental value of the mass. Up to now, the lowest mass of the vector tetraquark states  without strangeness obtained from the QCD sum rules is $4.24 \pm 0.10\,\rm{GeV}$ \cite{WZG-V-tetra-10}. While the $D\bar{D}_1$ molecular state with the $J^{PC}=1^{--}$ has a mass $4.36 \pm 0.08\,\rm{GeV}$, the lowest mass of the vector tetraquark molecular states from the QCD sum rules \cite{WZG-Y4260-mol-10-CPC}.

In Ref.\cite{WZG-Z4200-88-tetra-10}, we  construct the color-octet-color-octet type axialvector four-quark current to explore  the mass and
width of the $Z_c(4200)$ via the QCD sum rules, and obtain the  numerical values $M = 4.19 \pm 0.08\,\rm{GeV}$ and
$\Gamma \sim 334\,\rm{MeV}$, which  support assigning the $Z_c(4200)$ to be the color-octet-color-octet type tetraquark  state with the quantum numbers $J^{PC} = 1^{+-}$.
The $Z_c(4200)$ is observed in the $J/\psi \pi^+$ mass spectrum, while  the $Z_{cs}(4220)$ is observed in the $J/\psi K^+$ mass spectrum, it is natural to assign the $Z_{cs}(4220)$ to be the flavor $SU(3)$ cousin of the $Z_c(4200)$, i.e.  an axialvector color-octet-color-octet  type tetraquark state with the quantum numbers $J^{PC}=1^{+-}$,  by assuming small $SU(3)$ mass-breaking effect.

We can search for those hidden-charm tetraquark molecular states with strange and with hidden-strange shown in Table \ref{Assignments-Table} in  the $\eta_c K$, $\eta_c K^*$, $J/\psi K$, $J/\psi K^*$, $\eta_c\phi$, $J/\psi\phi$ invariant mass spectrum at the BESIII, LHCb, Belle II,  CEPC, FCC, ILC in the future, and confront the predicted masses to the experimental data, which maybe  shed light on the nature of the exotic $X$, $Y$, $Z$ states. On the other hand,
 we can take the pole residues $\lambda_Z$ as the basic input parameters to study the two-body
 strong decays of those hidden-charm tetraquark molecular states
with the  (light-cone) QCD sum rules, and obtain the partial decay widths (or the branching fractions)  to diagnose the nature of the exotic  states.

\begin{table}
\begin{center}
\begin{tabular}{|c|c|c|c|c|c|c|c|c|}\hline\hline
 $Z_c$($X_c$)                            & $J^{PC}$ & $T^2 (\rm{GeV}^2)$ & $\sqrt{s_0}(\rm GeV) $      &$\mu(\rm{GeV})$   &pole         &$|D(10)|$ \\ \hline

 $D^*\bar{D}^*$                          & $0^{++}$ & $2.8-3.2$          & $4.55\pm0.10$               &$1.6$             &$(40-62)\%$  &$\leq1\%$   \\

 $D^*\bar{D}_s^*$                        & $0^{++}$ & $2.9-3.3$          & $4.65\pm0.10$               &$1.6$             &$(41-63)\%$  &$<1\%$   \\

 $D_s^*\bar{D}_s^*$                      & $0^{++}$ & $3.1-3.5$          & $4.75\pm0.10$               &$1.6$             &$(40-61)\%$  &$\ll1\%$   \\ \hline

 $D\bar{D}^*-D^*\bar{D}$                  & $1^{++}$ & $2.7-3.1$          & $4.40\pm0.10$               &$1.3$             &$(40-63)\%$  &$\ll1\%$   \\

 $D\bar{D}_s^*-D^*\bar{D}_s$              & $1^{++}$ & $2.9-3.3$          & $4.55\pm0.10$               &$1.3$             &$(41-63)\%$  &$\ll1\%$   \\

 $D_s\bar{D}_s^*-D_s^*\bar{D}_s$          & $1^{++}$ & $3.0-3.4$          & $4.65\pm0.10$               &$1.3$             &$(42-63)\%$  &$\ll1\%$   \\ \hline

$D\bar{D}^*+D^*\bar{D}$                  & $1^{+-}$ & $2.7-3.1$          & $4.40\pm0.10$               &$1.3$             &$(40-63)\%$  &$\ll1\%$   \\

$D\bar{D}_s^*+D^*\bar{D}_s$              & $1^{+-}$ & $2.9-3.3$          & $4.55\pm0.10$               &$1.3$             &$(41-63)\%$  &$\ll1\%$   \\

$D_s\bar{D}_s^*+D_s^*\bar{D}_s$          & $1^{+-}$ & $3.0-3.4$          & $4.65\pm0.10$               &$1.3$             &$(42-63)\%$  &$\ll1\%$   \\  \hline

$D^*\bar{D}^*$                           & $1^{+-}$ & $3.0-3.4$          & $4.55\pm0.10$               &$1.6$             &$(42-63)\%$  &$<1\%$   \\

 $D^*\bar{D}_s^*$                        & $1^{+-}$ & $3.2-3.6$          & $4.65\pm0.10$               &$1.6$             &$(41-61)\%$  &$\ll1\%$   \\

 $D^*_s\bar{D}_s^*$                      & $1^{+-}$ & $3.3-3.7$          & $4.75\pm0.10$               &$1.6$             &$(42-61)\%$  &$\ll1\%$   \\ \hline

 $D^*\bar{D}^*$                          & $2^{++}$ & $3.0-3.4$          & $4.55\pm0.10$               &$1.6$             &$(41-62)\%$  &$<1\%$   \\

 $D^*\bar{D}_s^*$                        & $2^{++}$ & $3.2-3.6$          & $4.65\pm0.10$               &$1.6$             &$(40-60)\%$  &$\ll1\%$   \\

 $D_s^*\bar{D}_s^*$                      & $2^{++}$ & $3.3-3.7$          & $4.75\pm0.10$               &$1.6$             &$(41-61)\%$  &$\ll1\%$   \\

\hline\hline
\end{tabular}
\end{center}
\caption{ The Borel parameters, continuum threshold parameters, energy scales of the QCD spectral densities,  pole contributions, and the contributions of the vacuum condensates of dimension $10$  for the ground state hidden-charm tetraquark molecular states. }\label{BorelP}
\end{table}

\begin{table}
\begin{center}
\begin{tabular}{|c|c|c|c|c|c|c|c|c|}\hline\hline
$Z_c$($X_c$)                                                            & $J^{PC}$  & $M_Z (\rm{GeV})$   & $\lambda_Z (\rm{GeV}^5) $             \\ \hline

$D^*\bar{D}^*$                                                          & $0^{++}$  & $4.02\pm0.09$      & $(4.30\pm0.72)\times 10^{-2}$           \\

$D^*\bar{D}_s^*$                                                        & $0^{++}$  & $4.10\pm0.09$      & $(5.00\pm0.83)\times 10^{-2}$           \\

$D_s^*\bar{D}_s^*$                                                      & $0^{++}$  & $4.20\pm0.09$      & $(5.86\pm0.98)\times 10^{-2}$           \\ \hline

$D\bar{D}^*-D^*\bar{D}$                                                 & $1^{++}$  & $3.89\pm0.09$      & $(1.72\pm0.30)\times 10^{-2}$           \\

$D\bar{D}_s^*-D^*\bar{D}_s$                                             & $1^{++}$  & $3.99\pm0.09$      & $(1.96\pm0.35)\times 10^{-2}$           \\

$D_s\bar{D}_s^*-D_s^*\bar{D}_s$                                         & $1^{++}$  & $4.07\pm0.09$      & $(2.07\pm0.37)\times 10^{-2}$           \\ \hline

$D\bar{D}^*+D^*\bar{D}$                                                 & $1^{+-}$  & $3.89\pm0.09$      & $(1.72\pm0.30)\times 10^{-2}$           \\

$D\bar{D}_s^*+D^*\bar{D}_s$                                             & $1^{+-}$  & $3.99\pm0.09$      & $(1.96\pm0.35)\times 10^{-2}$           \\

$D_s\bar{D}_s^*+D_s^*\bar{D}_s$                                         & $1^{+-}$  & $4.07\pm0.09$      & $(2.07\pm0.37)\times 10^{-2}$           \\  \hline

$D^*\bar{D}^*$                                                          & $1^{+-}$  & $4.02\pm0.09$      & $(2.33\pm0.35)\times 10^{-2}$           \\

$D^*\bar{D}_s^*$                                                        & $1^{+-}$  & $4.11\pm0.09$      & $(2.71\pm0.41)\times 10^{-2}$           \\

$D_s^*\bar{D}_s^*$                                                      & $1^{+-}$  & $4.19\pm0.09$      & $(3.12\pm0.47)\times 10^{-2}$           \\ \hline

$D^*\bar{D}^*$                                                          & $2^{++}$  & $4.02\pm0.09$      & $(3.29\pm0.51)\times 10^{-2}$           \\

$D^*\bar{D}_s^*$                                                        & $2^{++}$  & $4.11\pm0.09$      & $(3.84\pm0.59)\times 10^{-2}$           \\

$D_s^*\bar{D}_s^*$                                                      & $2^{++}$  & $4.19\pm0.09$      & $(4.42\pm0.67)\times 10^{-2}$           \\
\hline\hline
\end{tabular}
\end{center}
\caption{ The masses and pole residues of the ground state hidden-charm tetraquark molecular states. }\label{mass-Table}
\end{table}

\begin{table}
\begin{center}
\begin{tabular}{|c|c|c|c|c|c|c|c|c|}\hline\hline
$Z_c$($X_c$)                                                            & $J^{PC}$   & $M_Z (\rm{GeV})$   & Assignments              \\ \hline

$D^*\bar{D}^*$                                                          & $0^{++}$   & $4.02\pm0.09$      &             \\

$D^*\bar{D}_s^*$                                                        & $0^{++}$   & $4.10\pm0.09$      &              \\

$D_s^*\bar{D}_s^*$                                                      & $0^{++}$   & $4.20\pm0.09$      &             \\ \hline

$D\bar{D}^*-D^*\bar{D}$                                                 & $1^{++}$   & $3.89\pm0.09$       & ? $X_c(3872)$           \\

$D\bar{D}_s^*-D^*\bar{D}_s$                                             & $1^{++}$   & $3.99\pm0.09$       &              \\

$D_s\bar{D}_s^*-D_s^*\bar{D}_s$                                         & $1^{++}$   & $4.07\pm0.09$       &             \\ \hline

$D\bar{D}^*+D^*\bar{D}$                                                 & $1^{+-}$   & $3.89\pm0.09$       & ? $Z_c(3900)$             \\

$D\bar{D}_s^*+D^*\bar{D}_s$                                             & $1^{+-}$   & $3.99\pm0.09$       & ? $Z_{cs}(3985/4000)$             \\

$D_s\bar{D}_s^*+D_s^*\bar{D}_s$                                         & $1^{+-}$   & $4.07\pm0.09$       &               \\  \hline

$D^*\bar{D}^*$                                                          & $1^{+-}$   & $4.02\pm0.09$       & ? $Z_c(4020)$             \\

$D^*\bar{D}_s^*$                                                        & $1^{+-}$   & $4.11\pm0.09$       & ?? $Z_{cs}(4100)$      \\

$D_s^*\bar{D}_s^*$                                                      & $1^{+-}$   & $4.19\pm0.09$       &             \\ \hline

$D^*\bar{D}^*$                                                          & $2^{++}$   & $4.02\pm0.09$       &                \\

$D^*\bar{D}_s^*$                                                        & $2^{++}$   & $4.11\pm0.
09$       &              \\

$D_s^*\bar{D}_s^*$                                                      & $2^{++}$   & $4.19\pm0.09$       &               \\
\hline\hline
\end{tabular}
\end{center}
\caption{ The possible assignments of the ground state hidden-charm tetraquark molecular states, the isospin limit is implied. We use two ?? to denote that
there is only a weak hint for the $Z_{cs}(4100)$ in the BESIII data. }\label{Assignments-Table}
\end{table}

\begin{table}
\begin{center}
\begin{tabular}{|c|c|c|c|c|c|c|c|c|}\hline\hline
$Z_c$($X_c$)                                                            & $J^{PC}$  & $M_Z (\rm{GeV})$   & Assignments        &$Z_c^\prime$ ($X_c^\prime$)      \\ \hline

$[uc]_{S}[\overline{dc}]_{S}$                                           & $0^{++}$  & $3.88\pm0.09$      & ?\,$X(3860)$       &       \\

$[uc]_{A}[\overline{dc}]_{A}$                                           & $0^{++}$  & $3.95\pm0.09$      & ?\,$X(3915)$       & \\

$[uc]_{\tilde{A}}[\overline{dc}]_{\tilde{A}}$                           & $0^{++}$  & $3.98\pm0.08$      &                    & \\

$[uc]_{V}[\overline{dc}]_{V}$                                           & $0^{++}$  & $4.65\pm0.09$      &                    & \\

$[uc]_{\tilde{V}}[\overline{dc}]_{\tilde{V}}$                           & $0^{++}$  & $5.35\pm0.09$      &                    &  \\

$[uc]_{P}[\overline{dc}]_{P}$                                           & $0^{++}$  & $5.49\pm0.09$      &                    &  \\ \hline

$[uc]_S[\overline{dc}]_{A}-[uc]_{A}[\overline{dc}]_S$                   & $1^{+-}$  & $3.90\pm0.08$      & ?\,$Z_c(3900)$      &?\,$Z_c(4430)$    \\

$[uc]_{A}[\overline{dc}]_{A}$                                           & $1^{+-}$  & $4.02\pm0.09$      & ?\,$Z_c(4020/4055)$ &?\,$Z_c(4600)$        \\

$[uc]_S[\overline{dc}]_{\widetilde{A}}-[uc]_{\widetilde{A}}[\overline{dc}]_S$     & $1^{+-}$   & $4.01\pm0.09$    & ?\,$Z_c(4020/4055)$ &?\,$Z_c(4600)$     \\

$[uc]_{\widetilde{A}}[\overline{dc}]_{A}-[uc]_{A}[\overline{dc}]_{\widetilde{A}}$ & $1^{+-}$   & $4.02\pm0.09$    & ?\,$Z_c(4020/4055)$ &?\,$Z_c(4600)$    \\

$[uc]_{\widetilde{V}}[\overline{dc}]_{V}+[uc]_{V}[\overline{dc}]_{\widetilde{V}}$ & $1^{+-}$   & $4.66\pm0.10$    & ?\,$Z_c(4600)$      &    \\

$[uc]_{V}[\overline{dc}]_{V}$                                           & $1^{+-}$  & $5.46\pm0.09$      &                    &  \\

$[uc]_P[\overline{dc}]_{V}+[uc]_{V}[\overline{dc}]_P$                   & $1^{+-}$  & $5.45\pm0.09$      &                    &  \\
\hline

$[uc]_S[\overline{dc}]_{A}+[uc]_{A}[\overline{dc}]_S$                   & $1^{++}$  & $3.91\pm0.08$      & ?\,$X(3872)$       &   \\

$[uc]_S[\overline{dc}]_{\widetilde{A}}+[uc]_{\widetilde{A}}[\overline{dc}]_S$     & $1^{++}$   & $4.02\pm0.09$    &?\,$Z_c(4050)$ &   \\

$[uc]_{\widetilde{V}}[\overline{dc}]_{V}-[uc]_{V}[\overline{dc}]_{\widetilde{V}}$ & $1^{++}$   & $4.08\pm0.09$    &?\,$Z_c(4050)$ &    \\

$[uc]_{\widetilde{A}}[\overline{dc}]_{A}+[uc]_{A}[\overline{dc}]_{\widetilde{A}}$ & $1^{++}$   & $5.19\pm0.09$    &               & \\

$[uc]_P[\overline{dc}]_{V}-[uc]_{V}[\overline{dc}]_P$                   & $1^{++}$  & $5.46\pm0.09$      &                    &  \\
\hline

$[uc]_{A}[\overline{dc}]_{A}$                                           & $2^{++}$  & $4.08\pm0.09$      &?\,$Z_c(4050)$      & \\

$[uc]_{V}[\overline{dc}]_{V}$                                           & $2^{++}$  & $5.40\pm0.09$      &                    & \\
\hline\hline
\end{tabular}
\end{center}
\caption{ The possible assignments of the ground state hidden-charm tetraquark states, the isospin limit is implied \cite{WZG-hidden-cc-tetra-10}. }\label{Assignments-Table-tetraquark}
\end{table}

\begin{table}
\begin{center}
\begin{tabular}{|c|c|c|c|c|c|c|c|c|}\hline\hline
$Z_c$($X_c$)                                                            & $J^{PC}$  & $M_Z (\rm{GeV})$   & Assignments          \\ \hline

$[uc]_{S}[\overline{sc}]_{S}$                                           & $0^{++}$  & $3.97\pm0.09$      &                           \\

$[uc]_{A}[\overline{sc}]_{A}$                                           & $0^{++}$  & $4.04\pm0.09$      &                     \\

$[uc]_{\tilde{A}}[\overline{sc}]_{\tilde{A}}$                           & $0^{++}$  & $4.07\pm0.08$      &                     \\

$[uc]_{V}[\overline{sc}]_{V}$                                           & $0^{++}$  & $4.74\pm0.09$      &                     \\

$[uc]_{\tilde{V}}[\overline{sc}]_{\tilde{V}}$                           & $0^{++}$  & $5.44\pm0.09$      &                      \\

$[uc]_{P}[\overline{sc}]_{P}$                                           & $0^{++}$  & $5.58\pm0.09$      &                      \\ \hline

$[uc]_S[\overline{sc}]_{A}-[uc]_{A}[\overline{sc}]_S$                   & $1^{+-}$  & $3.99\pm0.09$      & ?\,$Z_{cs}(3985/4000)$        \\

$[uc]_{A}[\overline{sc}]_{A}$                                           & $1^{+-}$  & $4.11\pm0.09$      &         \\

$[uc]_S[\overline{sc}]_{\widetilde{A}}-[uc]_{\widetilde{A}}[\overline{sc}]_S$     & $1^{+-}$   & $4.10\pm0.09$    &     \\

$[uc]_{\widetilde{A}}[\overline{sc}]_{A}-[uc]_{A}[\overline{sc}]_{\widetilde{A}}$ & $1^{+-}$   & $4.11\pm0.09$    &      \\

$[uc]_{\widetilde{V}}[\overline{sc}]_{V}+[uc]_{V}[\overline{sc}]_{\widetilde{V}}$ & $1^{+-}$   & $4.75\pm0.10$    &      \\

$[uc]_{V}[\overline{sc}]_{V}$                                           & $1^{+-}$  & $5.55\pm0.09$      &                      \\

$[uc]_P[\overline{sc}]_{V}+[uc]_{V}[\overline{sc}]_P$                   & $1^{+-}$  & $5.54\pm0.09$      &                     \\
\hline

$[uc]_S[\overline{sc}]_{A}+[uc]_{A}[\overline{sc}]_S$                   & $1^{++}$  & $3.99\pm0.09$      &          \\

$[uc]_S[\overline{sc}]_{\widetilde{A}}+[uc]_{\widetilde{A}}[\overline{sc}]_S$     & $1^{++}$   & $4.11\pm0.09$    &    \\

$[uc]_{\widetilde{V}}[\overline{sc}]_{V}-[uc]_{V}[\overline{sc}]_{\widetilde{V}}$ & $1^{++}$   & $4.17\pm0.09$    &     \\

$[uc]_{\widetilde{A}}[\overline{sc}]_{A}+[uc]_{A}[\overline{sc}]_{\widetilde{A}}$ & $1^{++}$   & $5.28\pm0.09$    &                \\

$[uc]_P[\overline{sc}]_{V}-[uc]_{V}[\overline{sc}]_P$                   & $1^{++}$  & $5.55\pm0.09$      &                      \\
\hline

$[uc]_{A}[\overline{dc}]_{A}$                                           & $2^{++}$  & $4.17\pm0.09$      & \\

$[uc]_{V}[\overline{dc}]_{V}$                                           & $2^{++}$  & $5.49\pm0.09$      &                  \\
\hline\hline
\end{tabular}
\end{center}
\caption{ The possible assignments of the ground state hidden-charm tetraquark states with strangeness \cite{WZG-Zcs3985-tetra-10}. }\label{Assignments-Table-tetraquark-Zcs}
\end{table}

\section{Conclusion}
In this article,  we  study the  mass spectrum of the ground state hidden-charm tetraquark molecular states  without strange, with strange and with hidden-strange
via the QCD sum rules in a comprehensive way by  carrying out the operator product expansion up to the vacuum condensates of dimension $10$ in a consistent way.
We take  account of the light flavor $SU(3)$ breaking effects in a consistent way and use the modified energy scale formula to determine the ideal energy scales of the QCD spectral densities.  Finally, we obtain the mass spectrum of the ground state hidden-charm tetraquark molecular states, which can be confronted to the experimental data in the future at  the BESIII, LHCb, Belle II,  CEPC, FCC, ILC by investigating  the invariant mass spectrum of the $\eta_c K$, $\eta_c K^*$, $J/\psi K$, $J/\psi K^*$, $\eta_c\phi$, $J/\psi\phi$.
   We can reproduce the experimental values of the masses of the  $X_c(3872)$, $Z_c(3900)$, $Z_{cs}(3985/4000)$ and $Z_c(4020)$ both in the  scenario of tetraquark  states and tetraquark molecular states.  Those exotic  $X$, $Y$ and $Z$ states maybe have a diquark-antidiquark type tetraquark core with the typical size of the $q\bar{q}$-type  mesons, the strong couplings  to the meson-meson pairs   lead to some
tetraquark molecule Fock components, and they maybe spend  a rather large time as the  tetraquark   molecular states, just like the $f_0(980)$ and $a_0(980)$. Or there maybe exist double exotic states with the almost  degenerated masses besides the same quantum numbers, just like the $Z_c(3900)$ and $Z_c(3885)$.

\section*{Acknowledgements}
This  work is supported by National Natural Science Foundation, Grant Number  11775079.

\end{document}